\documentclass{LT23auth}
\usepackage{graphicx}

\begin{document}

\begin{frontmatter}

\title{Finite-size effects on the thermal conductivity of $\rm^4$He near T$\rm_\lambda$}

\author[address1]{Michael T\"opler},
\author[address1]{Volker Dohm}

\address[address1]{Institut f\"ur Theoretische Physik,
Technische Hochschule Aachen, D-52056 Aachen, Germany}

\begin{abstract}
We present results of a renormalization-group calculation of the thermal
conductivity of confined $\rm^4$He in a $L^2 \times \infty$ geometry above and at
$T_\lambda$ within model F with Dirichlet boundary conditions for the order
parameter. We assume a heat flow parallel to the boundaries which implies
Neumann boundary conditions for the entropy density.
No adjustable parameters other than those known from bulk theory
and static finite-size theory are used.
Our theoretical results are compared with experimental data by Kahn and Ahlers.
\end{abstract}

\begin{keyword}
critical phenomena; finite-size effects; helium4; thermal
conductivity
\end{keyword}
\end{frontmatter}

\section{Introduction}

The theory of finite-size effects near phase transitions is an
area of active research. Well suited for a comparison
between theory and experiment is the superfluid transition of
$^4$He \cite{1}. So far, however, primarily {\it static}
properties have been investigated. Very little is known about
finite size effects on {\it dynamic} quantities. In particular no
theoretical prediction exists for the validity of dynamic
finite-size scaling along the lambda line of $^4$He. Here we present
some results of the first renormalization-group calculation of
finite-size effects on the thermal conductivity of $^4$He above and
at $T_\lambda$ and compare our results with experimental data by Kahn
and Ahlers \cite{2}.

\section{Theory}

Our calculations are based on model F \cite{3} which is defined by

\begin{equation}
\dot{\psi}_0 = - 2 \Gamma_0 \frac{\delta H}{\delta \psi^*_0} + i g_0
\psi_0 \frac{\delta H}{\delta m_0} + \Theta_\psi \; ,
\end{equation}

\begin{equation}
\dot{m}_0 = \lambda_0 \nabla^2 \frac{\delta H}{\delta m_0} + g_0
\nabla {\bf j}_s^0 + W_0 + \Theta_m \; ,
\end{equation}

\begin{eqnarray}
H &=& \int d^3 {\bf r} (\frac{1}{2} r_0 | \psi_0 |^2 + \frac{1}{2} |
\nabla \psi_0 |^2 + \tilde u_0 | \psi_0 |^4\cr
&&+ \frac{1}{2} \chi_0^{-1} m_0^2 + \gamma_0 m_0 | \psi_0 |^2 ) \;,
\end{eqnarray}

where

\begin{equation}
{\bf j}_s^0 ({\bf r}, t) \equiv {\rm Im} (\psi_0^* ({\bf r}, t) \nabla
\psi_0 ({\bf r}, t)) \;,
\end{equation}

\begin{equation}
r_0 = r_{0c} + a_0 \tilde t \; , \tilde t = (T - T_\lambda) / T_\lambda \; .
\end{equation}

We consider a rectangular $L^2 \times \widetilde L$ box geometry and assume
a stationary heat current {\it Q} in the {\it z} direction which
is generated by a heat source in the bottom plane $z = -
\widetilde L / 2$ and absorbed by a sink in the top plane $z =
\widetilde L/2$,

\begin{equation}
W_0 ({\bf r}) = Q [\delta (z + \widetilde L/2) - \delta (z -
\widetilde L/2)] \;.
\end{equation}

We impose Dirichlet boundary conditions for the order parameter
($\psi_0 = 0$) and Neumann boundary conditions for the entropy
density $m_0$ (vanishing spatial derivatives perpendicular to the
sidewalls). Eventually we let $\widetilde L \to \infty$ and
define the superfluid current ${\bf j} = \lim_{\widetilde L
\to \infty}  < {\bf j}_s^0 >$ in the stationary state. We are
interested in the finite-size effect on the thermal conductivity

\begin{equation}
\lambda_T (\tilde t, L) = \lambda_0 \left[1 + \frac{g_0}{L^2}
\lim_{Q \to 0} \frac{\partial}{\partial Q} \int\limits_{L^2} dx \;
dy j_z \right]^{-1}
\end{equation}

where $j_z$ is the $z$ component of ${\bf j}$. We have calculated
$\lambda_T (\tilde t, L)$ analytically for $\tilde t \geq 0$ up to
one-loop order employing the minimal renormalization approach at
fixed dimension $d = 3$ \cite{4} and using the effective
parameters known from bulk theory \cite{5}. We neglect possible
non-scaling contributions due to cutoff effects and van der Waals
forces \cite{6}.

\section{Results}

In Fig. 1 we present our prediction of $\lambda_T (0, L)$ at
saturated vapor pressure. It agrees reasonably well with the
experimental result for holes $2 \mu m$ in diameter
\cite{2}.

\begin{figure}[tph]
\bigskip
\begin{center}\leavevmode
\includegraphics[width=0.9\linewidth]{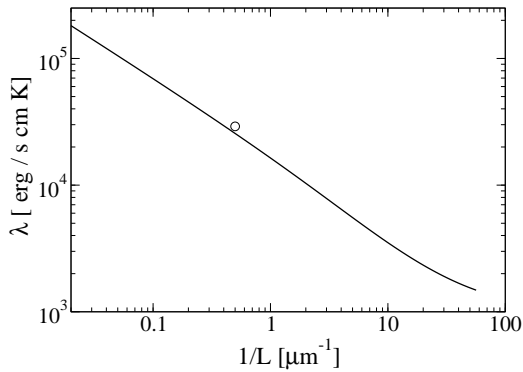}
\caption{Theoretical prediction for the thermal conductivity,
Eq. (7), at $T = T_\lambda$ as a function of $L^{-1}$. The
circle represents the experimental result by Kahn and Ahlers \cite{2} for
holes $2 \mu m$ in diameter.}
\label{Leitf_Tc}\end{center}\end{figure}

In Fig. 2 we plot our prediction for the relative deviation

\begin{equation}
\Delta \lambda = \frac{\lambda_b (\tilde t) - \lambda_T (\tilde t,
L)} {\lambda_b (\tilde t)}
\end{equation}

from the bulk thermal conducivity $\lambda_b (\tilde t) \equiv
\lambda_T (\tilde t, \infty)$ at saturated vapor pressure for
$L = 2 \mu m$ in the regime $\xi \ll L$ (solid line) where $\xi$
is the bulk correlation length above $T_\lambda$. It agrees
reasonably well with the experimental data \cite{2} in this regime.

\begin{figure}[tph]
\bigskip
\begin{center}\leavevmode
\includegraphics[width=0.9\linewidth]{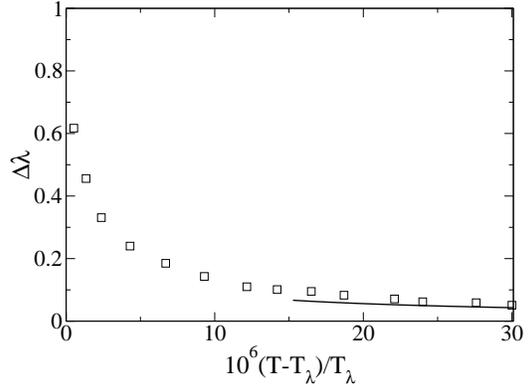}
\caption{Relative deviation $\Delta\lambda$, Eq. (8), versus
reduced temperature. Solid line: Theoretical prediction for
$L =2 \mu m$ in the regime $\xi \ll L$. Squares: Representative
set of data taken from Fig. 2 of Ref. \cite {2}.
}\label{Leitf_L}\end{center}\end{figure}

\newpage
Our analytical theoretical expression for $\lambda_T (\tilde t,
L)$ will be presented elsewhere \cite{7}. It will enable us to
make quantitative predictions for the finite-size scaling function
of $\lambda_T$ and for the pressure dependence of the
finite-size effects along the lambda line. This will provide the
basis for testing the range of validity of universal dynamic
finite-size scaling.

\begin{ack}

We acknowledge support by DLR and by NASA under grant numbers
50WM9911 and 1226553.

\end{ack}

\end{document}